\documentclass{article}
\usepackage{listings}
\usepackage{epsfig}
\usepackage{authblk}

\begin{document}

\title{Applying Type Oriented Programming to the PGAS Memory Model}
\author[1]{Nick Brown\footnote{Corresponding author: +44 (0) 131 650 6420, nick.brown@ed.ac.uk}}
\affil[1]{Edinburgh Parallel Computing Centre, James Clerk Maxwell Building, Kings Buildings, Edinburgh}
\date{}

\maketitle

\begin{abstract}
The Partitioned Global Address Space memory model has been popularised by a number of languages and applications. However this abstraction can often result in the programmer having to rely on some in built choices and with this implicit parallelism, with little assistance by the programmer, the scalability and performance of the code heavily depends on the compiler and choice of application.

We propose an approach, type oriented programming, where all aspects of parallelism are encoded via types and the type system. The type information associated by the programmer will determine, for instance, how an array is allocated, partitioned and distributed. With this rich, high level of information the compiler can generate an efficient target executable. If the programmer wishes to omit detailed type information then the compiler will rely on well documented and safe default behaviour which can be tuned at a later date with the addition of types. 

The type oriented parallel programming language Mesham, which follows the PGAS memory model, is presented. We illustrate how, if so wished, with the use of types one can tune all parameters and options associated with this PGAS model in a clean and consistent manner without rewriting large portions of code. An FFT case study is presented and considered both in terms of programmability and performance - the latter we demonstrate by a comparison with an existing FFT solver.
\end{abstract}

\section{Introduction}
As the problems that the HPC community looks to solve become more ambitious then the challenge will be to provide programmers, who might be non HPC experts, with usable and consistent abstractions which still allow for scalability and performance. Partitioned Global Address Space is a memory model providing one such abstraction and allows for the programmer to consider the entire system as one entire global memory space which is partitioned and each block local to some process. Numerous languages and frameworks exist to support this model but all, operating at this higher level, impose some choices and restrictions upon the programmer in the name of abstraction.

This paper proposes a trade-off between explicit parallelism, which can yield good performance and scalability if used correctly, and implicit parallelism which promotes simplicity and maintainability. Type oriented programming addresses the issue by providing the options to the end programmer to choose between explicit and implicit parallelism. The approach is to design new types, which can be combined to form the semantics of data governing parallelism. A programmer may choose to use these types or may choose not to use them and in the absence of type information the compiler will use a well-documented set of default behaviours. Additional type information can be used by the programmer to tune or specialise many aspects of their code which guides the compiler to optimise and generate the required parallelism code. In short these types for parallelisation are issued by the programmer to instruct the compiler to perform the expected actions during compilation and in code generation. They are predefined by expert HPC programmers in a type library and used by the application programmer who many not have specialist HPC knowledge.

Programmer imposed information about parallelism only appears in types at variable declaration and type coercions in expressions and assignments. A change of data partition or communication pattern only require a change of data types, while the traditional approaches may require rewriting the entire structure of the code. A parallel programming language, Mesham which follows the PGAS memory model, has been developed which follows this paradigm and we study a Fast Fourier Transformation (FFT) case study written in Mesham to evaluate the proposed approach. The pursuit for performance and scalability is a major objective of HPC and we compare the FFT Mesham version with that of an existing, mature solving framework and also consider issues of programmability. 

\section{The Challenge}
The difficulty of programming has been a challenge to parallel computing over the past several decades\cite{skillicorn}. Whilst numerous languages and models have been proposed, they mostly suffer from the same fundamental trade-off between simplicity and expressivity. Those languages which abstract the programmer sufficiently to allow for conceptual simplicity often far remove the programmer from the real world execution and impose upon them predefined choices such as the method of communication. The parallel programming solutions which provide the programmer with full control over their code often result in great amounts of complexity which can be difficult for even expert HPC programmers to master for non-trivial problems, let alone the non-expert scientific programmers which often require HPC.

PGAS languages, which provide for the programing memory model abstraction of a global address space which is partitioned and each portion local to a process also suffers from this trade off. For instance, to achieve this memory model the programmer operates at a higher level far removed from the actual hardware and often key aspects, such as the form of data communication, are abstracted away with the programmer having no control upon these key attributes. Operating in a high level environment, without control of lower level decisions, can greatly affect performance and scalability of codes with the programmer reliant on the compiler ``making the right choice'' when it comes to some critical aspects of parallelism.

Whilst the PGAS memory abstraction is a powerful one, on its own it still leaves complexity to the end programmer in many cases. For example changing the distribution of data amongst the processes can still require the programmer to change numerous aspects of their code. 

\section{Type oriented programming}

The concept of a type will be familiar to many programmers. A large subset of languages follow the syntax \emph{Type Variablename}, such as \emph{int a} or \emph{float b}, which is used to declare a variable. Such statements affect both the static and dynamic semantics - the compiler can perform analysis and optimisation (such as type checking) and at runtime the variable has a specific size and format. It can be thought that the programmer provides information, to the compiler, via the type. However, there is only so much that one single type can reveal, and so languages often include numerous keywords in order to allow for the programmer to specify additional information. Using the C programming language as an example, in order to declare a variable m to be a read only character where memory is allocated externally, the programmer writes \emph{extern const char m}. Where \emph{char} is the type and both \emph{extern} and \emph{const} are inbuilt language keywords. Whilst this approach works well for sequential languages, in the parallel programming domain there are potentially many more attributes which might need to be associated; such as where the data is located, how it is communicated and any restrictions placed upon this. Representing such a rich amount of information via multiple keywords would not only bloat the language, it might also introduce inconsistencies when keywords were used together with potentially conflicting behaviours.

Instead our approach is to allow for the programmer to encode all variable information via the type system, by combining different types together to form the overall meaning. For instance, \emph{extern const char m} becomes \emph{var m:Char::const::extern}, where \emph{var m} declares the variable, the operator \emph{:} specifies the type and the operator \emph{::} combines two types together. In this case, a \textbf{type chain} is formed by combining the types \emph{Char}, \emph{const} and \emph{extern}. Precedence is from right to left where, for example, the read only properties of the \emph{const} type override the default read \& write properties of \emph{Char}. It should be noted that some type coercions, such as \emph{Int::Char} are meaningless and so rules exist within each type to govern which combinations are allowed.

Within type oriented programming the majority of the language complexity is removed from the core language and instead resides within the type system. The types themselves contain their specific behaviour for different usages and situations. The programmer, by using and combining types, has a high degree of control which is relatively simple to express and modify. Not only this, the high level of type information provides a rich amount of information upon which the compiler can use and optimise the code. In the absence of detailed type information the compiler can apply sensible, well documented, default behaviour and the programmer can further specialise this using additional types if required at a later date. The result is that programmers can get their code running and then further tune if needed by using additional types. 

Benefits of writing type oriented parallel codes are as follows:
\begin{enumerate}
\item {\bf Simplicity} - by providing a well documented, clean, type library the programmer can easily control all aspects of parallelism via types or rely on default well-documented behaviour.
\item {\bf Efficiency} - due to the rich amount of high level information provided by the programmer the compiler can perform much optimisation upon the code. The behaviour of types can control the tricky, low level, details which are essential to performance and can be implemented by domain experts which are then used by non-expert parallel programmers.
\item {\bf Flexibility} - often initial choices made, such as the method of data decomposition, can retrospectively turn out to be inappropriate. However, if one is not careful these choices can be difficult to change once the code has matured. By using types the programmer can easily change fundamental aspects by modifying the type with the compiler taking care of the rest. At a language level, containing the majority of the language complexity in a loosely coupled type library means that adding, removing or modifying the behaviour of types has no language wide side effect and the ``core'' language is kept very simple.
\item {\bf Maintainability} - the maintainability of parallel code is essential. Current production parallel programs are often very complex and difficult to maintain. By providing for simplicity and flexibility it is relatively simple for the code to be modified at a later stage.
\end{enumerate}

\section{Mesham}

A parallel programming language, Mesham\cite{mesham}, has been created based around an imperative programming language with extensions to support the type oriented concept. By default the language follows the Partitioned Global Address Space memory model where the entire global memory, which is accessible from every process, is partitioned and each block has an affinity with a distinct process. Reading from and writing to memory (either local or another processes' chunk) is achieved via normal variable access and assignment. By default, in the absence of further types, communication is one sided but this can be overridden using optional additional type information.

The language itself has fifty types in the external type library. Around half of these are similar in scope to the types introduced in the previous section and other types are more complex allowing one to control aspects such as explicit communication, data composition and data partitioning \& distribution. In listing \ref{lst:dftOnesided} the programmer is allocating two integers, \emph{a} and \emph{b} on lines one and two respectively. They exist as a single copy in global memory and variable \emph{a} is held in the memory of process zero, \emph{b} is in the memory associated with process two. At line three the assignment (using operator \emph{:=} in Mesham) will copy the value held in b at process two into variable \emph{a} which resides in the memory of process zero. In the absence of any further type information the communication associated with such an assignment is one-sided, which is guaranteed to be safe and consistent but might not be particularly performant. 

\begin{lstlisting}[frame=lines,caption={Default one sided communication},label={lst:dftOnesided}]
var a : Int :: allocated[single[on[0]];
var b : Int :: allocated[single[on[2]]];
a:=b;
\end{lstlisting}

The code in listing \ref{lst:p2p} looks very similar to that of listing \ref{lst:dftOnesided} with one important modification, at line one the type \emph{channel} has been added into the type chain of variable \emph{a}. This type will create an explicit point to point communication link between process two and zero which means that any assignments involving variable \emph{a} between these processes will use the point to point link rather than one-sided. By default the \emph{channel} type is blocking and control flow will pause until the data has been received by the target process; the programmer could further specialise this to use asynchronous (non-blocking) communication by appending the \emph{async} type into variable \emph{a}'s type chain. In such, asynchronous, cases the semantics of the language is such that the programmer issues explicit synchronisation points, either targeted at a specific variable or all variables, where it is guaranteed that outstanding asynchronous communications will be completed. It can be seen that in the tuning discussed here the programmer, using additional type information, guides the compiler to override the default behaviour. This can be done retrospectively once their parallel code is working and allows one to tune certain aspects which might be crucial to performance or scalability.

\begin{lstlisting}[frame=lines, caption={Override communication to blocking point to point},label={lst:p2p}]
var a : Int :: allocated[single[on[0]] :: channel[2,0];
var b : Int :: allocated[single[on[2]]];
a:=b;
\end{lstlisting}

The code examples considered in this section demonstrate that, following the traditional PGAS memory model, using types one can either rely on the simple, safe and well documented default behaviour, or associate additional information and override the defaults as required. Types used to specialise the behaviour are themselves responsible for their specific actions. The benefit of this is that by keeping the majority of the language complexity in the types contained within a loosely coupled type library, it not only results in a much simpler ``core'' language but also experts can architect types which simply plug into the language.

\subsection{Comparison}

Unified Parallel C (UPC)\cite{upc} is an extension to C designed for parallelism and follows the PGAS memory model. It does this with the addition of language keywords, such as \emph{shared} to marked shared variables, and functions. Due to the limited nature of associating attributes to data using keywords there are still decisions which the UPC programmer is stuck with such as one-sided communication and the programmer is reliant upon the compiler to do the best job it can of optimisation in this regard. Additionally, whilst the memory model is global and communication abstracted, the programmer is still stuck with having to work with low level concepts such as pointers. As discussed, in the type oriented programming model, many additional attributes can be associated with variables by the programmer if the defaults are not suitable. All this type information supports a higher level view of the code because the types controls the behaviour of variables and allows for the elimination of many function calls which are common in more traditional approaches.

High Performance Fortran(HPF)\cite{hpf} is a parallel extension of Fortran90. The programmer specifies just the data partitioning and allocation, with the compiler responsible for the placement of computation and communication. The type oriented approach differs because programmer can, via types, control far more aspects of parallelism. Alternatively, if not provided, the type system allows for a number of defaults to be used instead. Co-array Fortran (CAF)\cite{caf} provides the programmer with a greater degree of control than in HPF, but still the method of communication is implicit and determined by the compiler whilst synchronisations are explicit. CAF uses syntactically shorthanded communication commands like Y[:]=X and synchronisation statements. Having these commands hard wired into the language is popular, not just with CAF but many other parallel languages too, the result is less flexible and more difficult to implement.

Titanium\cite{titanium} is a PGAS extension to the Java programming language. The PGAS memory model is followed as the implicit model but also allows the programmer to use explicit message passing constructs by using additional language facilities. In this respect, providing for both a higher level implicit memory model and more detailed explicit message passing model, Titanium has some similarities to Mesham. However explicit control in Titanium relies on the programmer issuing in built language keywords such as \emph{broadcast E from p} and/or object methods which results in language bloat. In Titanium moving from the default PGAS memory model to the more explicit message passing requires rewriting portions of the code, whereas with our approach the programmer just needs to modify the type which directs the compiler as to the appropriate way of handing communication. The Mesham type system is designed such that it allows the compiler to generate all possible communication options just by using additional types.

Chapel\cite{chapel} has been designed, similar to Mesham and Titanium, to allow the programmer to express different abstractions of parallelism. It does this by providing higher and lower levels of abstractions which support automating the common forms of parallel programming via the former and the optimisation and tuning of specific factors using the later. There are some critical differences between Mesham and Chapel. Firstly, many of these higher level constructs in Chapel, such as a reduction is implemented via an inbuilt operator, instead in Mesham these would be types in an independent library. In Chapel, if one declares a single data variable and then writes to it from multiple parallel processes at the same time then this can result in a race condition. The solution is to use a synchronisation variable, via the \emph{sync} keyword in the variables declaration. In the type based approach the Mesham programmer would be using a \emph{sync} type, instead of an inbuilt language keyword, one benefit of this is that if multiple synchronisation constructs were being used (such as Chapel's \emph{sync}, \emph{single} and \emph{atomic} keywords) then the behaviour in a type chain where precedence is from right to left is well defined. Whilst languages such as Chapel might disallow combinations of these keywords, supporting them in a type chain allows for the programmer to mix the behaviours of different synchronisations in a predicable manner which might be desirable.

\section{FFT case study}
FFTs are of critical importance to a wide variety of scientific applications ranging from digital signal processing to solving partial differential equations. Parallelised 2D Fast Fourier Transformation (FFT) code is far more complicated than the equivalent sequential code. Direct message passing programming requires the end programmer to handle every detail of parallelisation including writing the appropriate communication commands, synchronizations, and correct index expressions that delimit the range of every partitioned array slice. Whilst using the PGAS memory model can help abstract some of these details the programmer is reliant upon assumptions imposed, in the name of abstraction, which can be costly in terms of scalability and-or performance with other aspects such as the details of data transposition still needing to be considered. A small change of how the data is partitioned or distributed may result in code rewriting. Orienting parallelism around types, however, can relieve the end programmer from writing low level details of parallelisation if these can be derived from the type information in code.

\lstset{caption=2D parallel FFT Mesham code, label=lst:fft,frame=lines}
\begin{lstlisting}
var n:=8192;
var p:=processes() * 2;
var i,j;

var S : array[complex,n,n]::allocated[row[]::single[0]];
var A : array[complex,n,n]::allocated[row[]::horizontal[p]::single[evendist[]]];
var B : array[complex,n,n]::allocated[col[]::horizontal[p]::single[evendist[]]];
var C : array[complex,n,n]::allocated[row[]::vertical[p]::single[evendist[]]]::share[B];

var sins : array[complex,n/2]::allocated[multiple[]];
computeSin(sins);
proc 0 {readfile(S, "image.dat")};

A:=S;

for j from 0 to A.localblocks - 1 {
	var bid:=A.localblockid[j];
	for i from A[bid].low to A[bid].high FFT(A[bid][i - A[bid].low],sins);
};

B:=A;

for j from 0 to C.localblocks - 1 {
	var bid:=C.localblockid[j];
	for i from C[bid].low to C[bid].high FFT(C[bid][i-C[bid].low],sins);	
};

S:=C;
proc 0 {writefile(S, "image.dat")};
\end{lstlisting}

Listing \ref{lst:fft} is the parallel aspects of the 2D FFT case study implemented in Mesham. For brevity the actual FFT computation algorithm, a Cooley–Tukey implementation, and other miscellaneous functions have been omitted. At line 5 the two dimensional array \emph{S} is declared to comprise of complex numbers be of size \emph{n} in each dimension, allocated row major fashion and a single copy of it resides upon process zero. This array is used to hold the initial data, an image which is read in at line 12 by process zero and then the results of the transform are placed into it and written back out at line 29. Line 6 declares variable \emph{A}, again \emph{n} by \emph{n} complex numbers, but this time it is partitioned via the \emph{horizontal} type into \emph{p} distinct partitions which are evenly distributed amongst the processes using the \emph{evendist} type. This even distribution follows a cyclical approach where partitioned blocks will be allocated to process after process and can cycle around if there are more blocks than processes. Line 7 declares the 2D array \emph{B} to be sized, partitioned and distributed in a similar manner to that of \emph{A} but this array is indexed column major. The last partitioned array to be declared,\emph{C} which uses vertical partitioning rather than horizontal, shares the underlying memory with \emph{B}; in effect this is a different view or abstraction of some existing memory.

Line 10 declares the sinusoid array. Using the \emph{multiple} type without further information results in allocation to the memory of all processes and this is used to compute the pre-calculated constant sinusoid parameters needed by the FFT kernel. Note that in this case no explicit array ordering is provided, in the absence of further information arrays default to row major ordering. In fact we could have omitted all \emph{row} types in the code if we had wished but these are provided to make explicit to the reader how the partitioned data is allocated and viewed. 

The assignment \emph{A:=S} at line 14 will result in a scattering of data held in \emph{S}, which is located on process zero, amongst the processes into each partitioned block of \emph{A}. In the loop at lines 16 to 19, each process will iterate through the blocks allocated to them and for each block perform the 1D FFT on individual rows. Assignment from \emph{A} to \emph{B} at line 21 essentially transposes \emph{A} and shuffles the blocks of array \emph{A} across processes. This allows each process to perform linear FFT on the other dimension locally. Because \emph{C} uses vertical partitioning and is a row major view of the data, performing row-wise FFT on \emph{C} is the same as performing column-wise FFT on \emph{B} at lines 23 to 26. The last assignment \emph{S:=C} gathers the data distributed amongst the processes into array \emph{S} held on process zero.

From the code listing it can be seen that the number of partitioned data blocks is two times the number of processes. Uneven partition sizes, for instance when the number of partitions does not divide evenly into the data size is transparent to the programmer. The types also abstract how and where the data is decomposed and processes can hold any number of blocks with the allocation, communication and transposition all taken care of by the type library. In conventional languages and frameworks it can add considerable complexity when blocks of data are uneven sizes and unevenly distributed, but using the type oriented approach this is all handled automatically. The programmer need not worry about these low level and tricky aspects - unless they want to where additional type information can be used to override the default behaviour.

\subsection{Modifying data decomposition and distribution}

It is often the case that programmers wish to get their parallel codes working in the first instance and then further tune and specialise if required. Often decisions made early on, such as the method of data decomposition, might not be correct retrospectively but can be very difficult to change without rewriting large portions of the code. Conversely, when orientating the code around types, changing the method of data decomposition is as simple as modifying a type. This will abstract exactly what data is where and allows for the programmer to not only tune but also experiment with different distribution options and how these can affect their code performance and scalability.

In listing \ref{lst:fft} the \emph{evendist} type has been used to perform an even cyclical distribution of the data. Instead, the programmer can change one or more of the distribution mechanisms to another distribution type such as array distribution. The \emph{arraydist} type allows the programmer to explicitly specify what blocks reside in the memory of what processes using an integer array. The index of each element in the array corresponds to the block Id and the value held there which process it resides upon. Listing \ref{lst:fftarraydist} illustrates using array distribution and is a snippet of the Mesham FFT code declaring the distributed arrays. At line 1 the array \emph{d} is declared to be an array of \emph{p} integers and in the absence of further information a copy of this is, by default, allocated on all processes. At lines 3 to 5 for every even numbered block Id we are allocating it to process one and uneven block Ids to process two. The arrays \emph{A}, \emph{B} and \emph{C} are then declared to use the \emph{arraydist} type with the array \emph{d} controlling what blocks belong where. Apart from modifying the type and code for the distribution, all other aspects of the FFT code in listing \ref{lst:fft} remain unchanged and the programmer can explicitly change what blocks belong where by modifying the values of the distribution array \emph{d}.

\lstset{caption=Mesham FFT example using array based data distribution, label=lst:fftarraydist,frame=lines}
\begin{lstlisting}
var d : array[Int,p];
var i;
for i from 0 to p - 1 {
	d[i]:=i % 2 == 0 ? 1 : 2;
};

var A : array[complex,n,n]::allocated[row[]::horizontal[p]::single[arraydist[d]]];
var B : array[complex,n,n]::allocated[col[]::horizontal[p]::single[arraydist[d]]];
var C : array[complex,n,n]::allocated[row[]::vertical[p]::single[arraydist[d]]]::share[B];

\end{lstlisting}
\subsection{Results}
\label{resultssection}
Whilst the programmability benefits of orienting parallel codes around types have been argued, it is equally important to consider the performance and scalability characteristics of this programming model. We have tested the Mesham version in code listing \ref{lst:fft}, which uses a Cooley–Tukey FFT kernel against the Fastest Fourier Transformation in the West version 3 (FFTW3)\cite{fftw} library. FFTW is a very commonly used and mature FFT calculation framework which looks to optimise the computational aspect of FFT by selecting the most appropriate solver kernel based upon parameters of the data. Performance testing has been carried out on HECToR, the UK National Supercomputer, a Cray XE6 with 32 cores per node, 32GB RAM per node and interconnection via the Gemini router. Data distribution in both test codes is that of even, cyclical, distribution with one block of data per process. The results presented in this section are the average of three runs.
 
\begin{center} 
\begin{figure}[htb]
\begin{center}\includegraphics{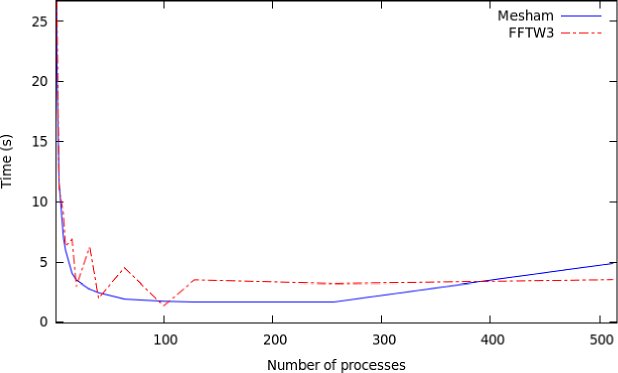}\end{center}
 \caption{Performance of Mesham FFT version compared to FFTW3}
    \label{fig:performance}
\end{figure}
\end{center}

Figure \ref{fig:performance} illustrates the performance of the FFT example in Mesham compared with the same problem solved using FFTW3. It can be seen that on small numbers of processes the performance is very similar and both exhibit good scalability as the number of cores is increased initially. There is some instability with the FFTW3 version compared to running the code using an even and uneven partitioning of data. Previous tests using FFTW2 illustrated that that older version of the library performed poorly when run parallel with uneven block sizes of data. Ironically in our tests the latest version, FFTW3, exhibits better performance when run with an uneven partitioning of data compared to an even partitioning. The performance of the Mesham version is more stable and predictable. The rich amount of information available at compile and runtime means that the language is able to select the most appropriate form of communication for specific situations automatically. The \emph{one size fits all} approach of communication adopted by many existing libraries is often optimised for specific cases and does not necessarily perform well in all configurations. At medium numbers of core counts the performance of the Mesham FFT version is more favourable than that of FFTW3 although as we go to larger numbers of processes the Mesham version does degrade faster. Due to the slightly larger overhead of the presently implemented Mesham parallel runtime system, performance degradation sets in somewhat earlier for this strong scaling scenario than in the highly tuned Cray MPI implementation.

Due to the abstractions provided by the PGAS memory model and our use of types, it is entirely possible to maintain correctness of the code whilst running on different architectures although this might have a performance impact. The implementation of Mesham is such that all architecture dependant aspects, for example how specific communications are implemented, are directed through a runtime abstraction layer which can be modified to suit different target machines. The runtime abstraction layer used for the experiments in this paper was for each PGAS processor to be single processes which are connected via MPI. A threading layer also exists which Mesham codes can use unmodified, and an avenue of further work will be to explore how we might optimise performance by selecting or mixing these layers. As previously noted, by changing types the programmer can very easily change key aspects of their code or experiment with different choices such as data decomposition, and this will promote easy tuning to specific architectures. Contrast against more traditional approaches, such as MPI, the porting of these codes to different architectures or mixing paradigms such as OpenMP with MPI often requires substantial and indepth changes to be made.

\subsection{Usage in library development}
The FFT case study that we have considered in listing \ref{lst:fft} simply illustrates the code in a single function. It is worth mentioning the suitability to more advanced codes, or even library development, where data using these complex type representations are passed between functions. In the current implementation of Mesham the entire type chain of a variable must be specified in the formal arguments of a function, which means that the compiler has detailed knowledge of the variables passed to a function and can perform appropriate static analysis and optimisations upon them. At runtime, when passed as an actual argument to a function, data will already have been allocated which occurs as part of a variable's declaration. The Mesham runtime library keeps track of the state of all program variables which means that during execution functions not only know the exact type of data but also its current state. The result is that, for the FFT example, no redistribution of the data would be required if passed to a function. 

\section{Conclusions}

This paper is not intended to describe the entire language Mesham but illustrate the central ideas behind the programming paradigm and demonstrate advantages when applied to the PGAS memory model. Aspects of this paradigm could, in the future, be used as part of existing PGAS languages to get the best of both worlds - a solution which parallel programmers are already familiar with but the added programmability benefits of our approach.

The rationale behind type oriented parallelism is not only to generate a highly efficient parallel executable but also enable programmers to write the source program in an intuitive and abstract style. The compiler essentially helps the programmer determine various sophisticated details of parallelisation as long as such details can be derived from the types in the source program. Optimization algorithms can also benefit from such additional type information. We have used a 2D parallel FFT case study to evaluate the success of our approach, both in terms of programmability with the benefits this affords, and also performance when compared to more traditional solving solutions. It has been seen how the Mesham programmer can architect their code at a high level using language default behaviour and then, by modifying type information, further specialise and tune whereas existing PGAS solutions often impose specific ``best effort" decisions upon the programmer. By using types programmers can even experiment with different choices, such as data decomposition, which traditionally require a much greater effort to modify.

We have compared the performance of the FFT Mesham case study against that of FFTW3. Whereas FFTW3 optimises heavily based upon the computation aspect; our version, where the compiler and runtime optimise the communication based upon the rich amount of type information, performs comparatively and in some instances favourably. There is further work to be done investigating why the performance of the Mesham version decreases more severely than FFTW past the optimal number of processes and we are looking to extend our version to 3D FFT with additional data decompositions such as Pencil. We also believe that Mesham would make a good platform for exploring heterogeneous PGAS, where the complexity of managing data stored on different devices can be abstracted via types. As discussed in section \ref{resultssection} all machine dependant aspects are current managed via a runtime abstraction layer, and further development of this could allow for existing codes to be run unmodified on these heterogeneous machines. 

\bibliography{document}
\bibliographystyle{plain}

\end{document}